\documentclass[%
 aip,
 amsmath,amssymb,
 reprint,%
]{revtex4-1}

\usepackage[utf8x]{inputenc}

\usepackage[T1]{fontenc}

\usepackage{amsfonts}

\usepackage{array}

\usepackage{multirow}
\usepackage{eurosym}

\usepackage{pslatex}
\usepackage{graphicx}
\usepackage{color}
\usepackage{dcolumn}

\begin{document}

\title{F{e}$^{3+}$ paramagnetic ion in $\alpha$-$Al_{2}O_{3}$ energy levels revisited. Application to a 31 GHz Maser proposal}
\author{M. Mrad, P.Y. Bourgeois, Y. Kersal\'e and V. Giordano}

\affiliation{FEMTO-ST Institute, Time and Frequency Dpt., 26 chemin de l'Epitaphe, 25030 Besan\c con Cedex, France}

\email{mohamad.mrad@femto-st.fr}

\date{\today}


\begin{abstract}

The molecular structure of a 3$d^{5}$ configuration ion in a trigonal ligand field is theoretically established on the basis of the 252$\times$252 complete energy matrice. The optical absorption and the electron paramagnetic resonance spectra of the $Fe^{3+}\!$ ion in the sapphire crystal ($\alpha\!\!-\!\!Al_{2}O_{3}$) have been studied by diagonalizing the complete energy matrice. The calculated results are in very good agreement with previous experimental observations. The strength of the transition probabilities between pairs of the energy levels have been calculated to determine the possibility to achieve a population inversion in the ground state by applying optical pumping to the  crystal. Preliminary results based on the computed transition parameters and on the maser rate equations model show that a 31 GHz maser signal can be effectively generated depending on the cryogenic resonator design.
\end{abstract}

\maketitle

\section{Introdution}

Masers based on paramagnetic ions hosted in a crystal were extensively studied and used as stable microwave sources or as low noise amplifiers in the 60's \cite{Sieg1964}. Maser action was obtained at low temperature in different  doped crystals, which the most popular was the sapphire doped with Cr$^{3+}$ or Rubis. Numerous types of maser were designed generally based on a doped crystal inserted into a waveguide or a metallic cavity cooled down the liquid Helium temperature. The doping level was typically of the order of 0.05\% or so and for a X-band source the cryogenic resonator presented a Q-factor of the order of 10,000 or less. Most of the systems were based on the ground state levels multiplicity to get the required population inversion through the action of a microwave pumping. Some attempts to optically pump the crystal have been realized \cite{gayda1971,devor1962}. But no real advantage compared to the classical maser scheme has been demonstrated. The lack of narrowband and easy to handle laser source as well as the poor quality of available doped crystals were certainly the main difficulties to achieve a reliable optically pumped microwave maser. Eventually the interest for this technology was almost stopped after the development of low noise figure cryogenic semiconductor amplifiers, stable quartz crystal oscillators and low noise frequency synthesis.  \\

More recently, we reported the possibility to get a 12 GHz maser signal presenting a short term  relative frequency instability below $1\times 10^{-14}$ at short term with a large marge of progress  \cite{Bour2005,Benm2007}. 
Such a frequency stability performance is required for a number of very demanding applications as metrology, space navigation, radioastronomy or fundamental physic experiments. The search for a reliable microwave source presenting a relative frequency stability of some $10^{-15}$ is a challenge that motivates a lot of innovative works all around the world \cite{gior-esa-2011, kessler2011, thorpe2011}.
In our first prototype the sustaining amplification is achieved in a cryogenic sapphire  whispering gallery mode resonator containing a small amount of paramagnetic $Fe^{3+}$ ions. The $Fe^{3+}$ ion exhibits three ground state energy levels. A population inversion is obtained between the two lower ground state levels by submitting the $Fe^{3+}$ ions to a 31 GHz signal. The maser oscillation takes place at 12.04 GHz. As the sapphire resonator exhibits at low temperature extremely low losses, the $Fe^{3+}$ ion concentration required to get enough gain is very low. Contrary to the maser of the 60's, our WhiGMO (Whispering Gallery Maser Oscillator) incorporates a microwave resonator presenting a Q-factor of typically one billion at 4.2K made from a sapphire crystal containing typically 0.1 ppm or less of active ions. The experimental validation of the 12 GHz WhiGMO and its potentiality to be competitive with the state-of-the-art microwave stable references opens the way to revisite the Maser concept and the associated material properties. Moreover the WhiGMO principle can be extented to other transitions of the $Fe^{3+}$ ion in sapphire or to other doped crystals to design stable sources at higher frequencies.

In this paper we refined the description of the $Fe^{3+}$ ion embedded into a ($\alpha$-$Al_{2}O_{3}$) matrix to derive its 252 energy levels with an improved accuracy compared to previous works \cite{Zhao1995}. Moreover we calculated the $Fe^{3+}$ ion strength transition probability including those for optical transitions. It appears than a few  allowed optical transitions are favorable to obtain a maser operation at 31 GHz applying optical pumping at wavelength achievable with available low cost laser sources. Eventually, a first preliminary evaluation of the 31 GHz maser concept is presented. It is shown that the 31 GHz maser signal power achievable is larger than the classical 12 GHz version, i.e. a power gain of 100 is expected in the optimized conditions. As fundamental limit of frequency stability is fixed by the thermal noise, any increase in the maser signal output will benefit to the performance.

\section{The molecular Hamiltonian}

The electronic configuration of the $Fe^{3+}$ cation is [Ar]3$d^{5}$. The spectroscopic term of the ground state is $^{6}S$ (S= 5/2 and L= 0). In a doped sapphire crystal the $Fe^{3+}$ ion replaces the cation $Al^{3+}$ and undergoes the action of the crystal field of the nearby ions $O^{2-}$ and $Al^{3+}$ (figure \ref{Fig disto}). The cation is located in a distorted octahedral. Assuming the cation $Fe^{3+}$ in symmetry $O_{h}$, the fundamental term will be $^{6}A_{1g}$($O_{h}$).

Since the appearance of the fundamental articles of Racah \cite{Racah1942}, Eliot and Stevens \cite{Elliot1953a} and Judd \cite{Judd1967}, the theoretical interpretation of electronic spectra of transition metals has became increasingly easy. The appropriate molecular Hamiltonian is:
\begin{align}
H&=H_{ee}+H_{so}+H_{c}+H_{Zeeman} \label{eq1} \\
&+ Trees \ correction + Racah \ correction \nonumber
\end{align}
In this section, we will discuss the electronic energy levels of a transition metal ion in crystalline solid. The theoretical foundation is already established and discussed in detail in the bibliography \cite{Judd1967, Judd1963, Wybo1965, Blea1986, Liu2005}. The Hamiltonian includes : the electrostatic Coulomb interactions ($H_{ee}$ ), the spin-orbit coupling ($H_{so}$ ),  the ion-ligand interactions described in the framework of crystal field theory ($H_{c}$), the Zeeman interaction, and the Trees and Racah corrections.

\subsection{Free ion}
Assuming the radial function of the  $Fe^{3+}$ d orbital as proposed by Zhao \cite{Zhao1994} using the "double-zeta exponential'' method, we find:
\begin{align}
R_{d}(r) & = 0.677(\frac{11.2^{7}}{6!})^{1/2} r^{2} exp(-5.6r) \nonumber\\
         & + 0.55237 (\frac{3.446^{7}}{6!})^{1/2} r^{2} exp(-1.732r)
\label{eq2}
\end{align}
From $R_{d}(r)$, the Racah parameters $B_{0}$ and $C_{0}$, the spin-orbit coupling constant $\zeta_{0}$, and the mean values $\langle r^{k}\rangle_{0}$ have been calculated for the free ion:
\begin{equation}
B_{0}= 1130.22 \ cm^{-1}\ \ ,\ \ C_{0}= 4111.45 \ cm^{-1} \nonumber
\end{equation}
\begin{equation}
\zeta_{0}= 588.946 \ cm^{-1}
\label{eq3}
\end{equation}
\begin{equation}
\langle r^{2}\rangle_{0}= 1.89039 \ a.u.\ \ ,\ \ \langle r^{4}\rangle_{0}= 11.46485 \ a.u. 
\nonumber
\end{equation}

We take into account the Trees \cite{Trees19} and the Racah \cite{Racah1951} corrections.
The constant of Trees correction $\alpha_{0}$= 40 $cm^{-1}$ and the Racah correction $\beta_{0}$= -131 $cm^{-1}$ are fitted from the spectra of the free ion \cite{Zhao1994}. 

\subsection{Crystal field interaction}

According Wybourne \cite{Wybo1965}, the potential of the crystal field can be written as a linear combination of irreducible tensor operators $C^{k}_{q}$:
\begin{equation}
H_{c}= \sum_{kq}{B_{kq}C^{k}_{q}}
\label{eq4}
\end{equation}
where $B_{kq}$ are the crystal field parameters. In the case of d-electrons, the allowed values for the integer $k$ are 0, 2, 4 and  $-k \leq q \leq +k$.

The structure of the equation~(\ref{eq4}) is determined by the symmetry of the crystal lattice. For a trigonal field ($C_{3v}$), the summation contains four terms: $B_{20}$, $B_{40}$, $B_{43}$ and $B_{4-3}$. In the generalized model of the crystal field, the trigonal crystal field parameters $B_{kq}$ were given by Zhao \cite{Zhao1995}. 
The values of the reduced matrix elements of tensor operators are published in the book of Nielson and Koster  \cite{Nielson1963} for 16$\times$16 matrices. From the Wigner-Eckart theorem we developped our generalized model based on 252$\times$252  matrices.

\subsection{Calculation method}

The general expression for the interaction between the central metal ion and ligand orbitals is complex, a reasonable approximation for the electrostatic parameters B and C, the constant of spin-orbit, and the average values $\langle r^{4}\rangle$ was made that :
\begin{align}
B=& N^{4}B_{0}\ \ \ \ ,\ \ \ \ \ C= N^{4}C_{0} \nonumber \\
\zeta=& N^{2}\zeta_{0}\ \ ,\ \  \langle r^{4}\rangle= N^{2}\langle r^{4}\rangle_{0} \nonumber\\
\alpha=& N^{4}\alpha_{0}\ \ \ \ ,\ \ \ \ \ \beta= N^{4}\beta_{0} \nonumber \\
\label{eq5}
\end{align}
where N is the average reduction factor due to the covalency.

\begin{figure}[h!]
\centering
\includegraphics[scale =0.6] {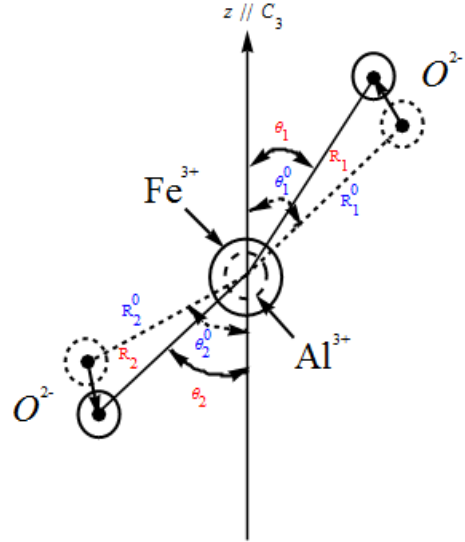}
\caption{Local structure distortion of octahedral $Fe^{3+}$ center in $\alpha$-$Al_{2}O_{3}$ system. \label{Fig disto}}
\end{figure}

In the $Al_{2}O_{3}$ crystal the position of $Al^{3+}$ is defined respectively to the sites of the two nearest cations as schematized in the figure \ref{Fig disto}. 

The intersite distances and the corresponding angles with respect the symetry axis $C_{3}$ have been determined in previous works  \cite{MacClure1963, Moorjani1963} as:

\begin{center}
\begin{align}
R_{1}^{0}= &1.966 \mathring{A} \ \ \ \ \ , \ \ \ \ \ \theta_{1}^{0}= 47.44^{o} \nonumber \\
R_{2}^{0}= &1.857 \mathring{A} \ \ \ \ \ , \ \ \ \ \ \theta_{2}^{0}= 62.70^{o}
\label{eq6}
\end{align}
\end{center}

To represent the crystal distortion due to the substitution of $Al^{3+}$ by $Fe^{3+}$, we assume as proposed by Zhao \cite{Zhao1998} that the two distances is affected by an identical relative variation. We add an identical constraint     to the angle relative variations, thus introducing two independent parameters $f$ and $g$ defined as 

\begin{equation}
f= \frac{R_{1}}{R_{1}^{0}}= \frac{R_{2}}{R_{2}^{0}}\ \ \ \ \ \ \ ; \ \ \ \ \ \ g= \frac{\theta_{1}}{\theta_{1}^{0}}= \frac{\theta_{2}}{\theta_{2}^{0}}
\label{eq7}
\end{equation}

Eventually the three independent parameters $N, f$ and $g$ have been varied to approach the experimental observations. We found the optical spectrum of $Fe^{3+}$ in  $\alpha-Al_{2}O_{3}$ is well represented by taking :

\begin{equation}
N= 0.863\ ,\ \ g= 0.996\ \ and\ \ f= 1.003
\label{eq8}
\end{equation}

The comparison between our calculations and experimental observations is shown in Table \ref{Tab Valuesa} compared to the values given by Zhao \cite{Zhao1995}, which the method serves as the basis of our own model. The introduction of a third independant adjustable parameter leads to an improvement in the determination of almost all the optical wavelengths in the $Fe^{3+}$ ion spectrum.

\begin{table}[h!!!]
\centering
\begin{tabular}{llll}
  \hline
$O_{h}$ &Zhao \cite{Zhao1995}  &  This work & Observed \cite{Lemann1970}\\
 \hline


               &   &   &  \\
$^{4}T_{1}(G)$ &  9790 & 9700 & 9450 - 9700 \\
               & 10048 & 9990 & \\
               & 10229 & 10128 &  \\

               & 10392 & 10294 & \\

               &  10473 &  10381 & \\

               &  10493 &  10396 &  \\

               &   &   &  \\

$^{4}T_{2}(G)$ &  14229 & 14183 & 14350  \\

               &  14262 & 14208 &  \\

               &  14315 &  14219 & \\

               &  14373 & 14256 &   \\

               &  14399 & 14281 &  \\

               &  14401 & 14282 & \\

             &   &   &  \\

$^{4}A_{1}(G)$ &   17977 & 17875 & 17600 - 17800 \\

               &   17998 & 17897 &  \\

$^{4}E(G)$     &  18101 & 18015 & \\

               &  18143 & 18056 & \\

               & 18188 & 18105 & \\

               &  18199 & 18113 & \\
              &   &   &  \\

$^{4}T_{2}(D)$ &  20181 & 20075 & \\

               & 20191 & 20078 & \\

               &   20341 & 20233 & \\

               &  20533 & 20414 & \\

               &  20562 & 20432 &  \\

               &  20823 &  20470 & \\

               &   &   &   \\

$^{4}E(D)$  &  22299 & 22195 & 22120 - 22200 \\

            &   22313 & 22212 &  \\

            & 22319 & 22219 &   \\

            &  22328 &  22229 &  \\

               &   &   &  \\
$^{4}T_{1}(P)$ &  25652 & 25663 &  25680 - 26700  \\

               & 25805 & 25755 & \\

               &  25945 & 25875 & \\

               & 26141 & 26090 &  \\

               &  26330 & 26225 & \\

               & 26457 & 26526 & \\

               &   &   &   \\

$^{4}A_{2}(F)$ &  30167 & 29884 & 29000  \\

               & 30170 & 29887 & \\
\hline
\end{tabular}
\caption{ Calculation and observed of the d-d transition for the crystal $\alpha$-$Al_{2}O_{3}:Fe^{3+}$ (the values are in $cm^{-1}$) .}
\label{Tab Valuesa}
\end{table}

To strengthen our confidence in our physical description, we derived from our model the spin-Hamiltonian parameters describing the ground state $^{6}A_{1}(S)$  of the $Fe^{3+}$ ion. Without applied static magnetic field, there are three energy levels labelled in the following $|i\rangle$ with $i=1,2$ or $3$ (see figure \ref{fig:Fe3+-ground-state}).
From the expression of spin-Hamiltonian given in \cite{Blea1953}, the energy difference between the three hyperfine levels of the ground state are written as :
\begin{equation}
E_{3}-E_{1}= \frac{1}{3}[(18D+a-F)^{2}+80a^{2}]^{1/2}  \label{eq9}\\
\end{equation}
\begin{equation}
E_{2}-E_{1}= \frac{3}{2}(a-F)-D+\frac{1}{6}[(18D+a-F)^{2}+80a^{2}]^{1/2}
\nonumber
\end{equation}

\begin{figure}[h!!!]
\centering
   \includegraphics[scale =0.3] {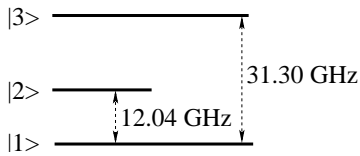}
\caption{Structure of $Fe^{3+}$ ground state \label{fig:Fe3+-ground-state} in the $\alpha$-$Al_{2}O_{3}$ crystal without applied static magnetic field. \label{fig:Fe3+-ground-state}}
\end{figure}

The calculated values of the spin-Hamiltonien parameters: D, a, a-F and the energy gaps are given in Table \ref{Tab ZFS} and compared to the previous evaluation due to Zhao and experimental values.

\begin{table}[h!!!!!!]
\centering
\begin{tabular}{c c c c}
  \hline  
 ZFS & Zhao 1995  &  This &  Observed\\
 parameters & \cite{Zhao1995} & article & \cite{Bog1959, Lee1977} \\
  \hline
 D  & 1784 & 1723 &  1718 $\pm$ 2 \\

a-F  & 322 & 330 & 329 $\pm$ 2 \\

 a  & 254 & 268 &   241 $\pm$ 4 \\

$E_{3}-E_{1}$  &  10838 & 10478 & 10451 \\ 

$E_{2}-E_{1}$ & 4118 & 4011 &  4015  \\
\hline
\end{tabular}
\caption{Calculated spin-Hamiltonian parameters for $\alpha$-$Al_{2}O_{3}:Fe^{3+}$ at 300 K compared to experimental values (the unit is $10^{-4}$ cm$^{-1}$).\label{Tab ZFS}}
\end{table}

\section{transition probability}

The optical spectrum of the iron doped sapphire shows fairly intense absorption bands \cite{Lemann1970,Lemann1971} indicating that we can interact optically with the $Fe^{3+}$ ions.
Our objective is to evaluate the possibility to get a population inversion in the ground state by applying an optical radiation to the crystal. Thus we need to calculate the effect of an optical radiation on the population of each ground state level $|i\rangle$, with $i=1,2 \mathrm{~or~}3$ taking into account the allowed transitions with the different excited levels $|j\rangle$ listed in the table \ref{Tab Valuesa}.  Following Siegman  \cite{Sieg1964}  the transition rate $W_{ij}$ between one ground state level $|i\rangle$ and one excited level $|j\rangle$ is given by:
\begin{equation}
W_{ij}= \frac{1}{4}(\gamma \mathrm{H})^{2} g(\nu) \sigma^{2}
\label{equ:taux-transition}
\end{equation}
Where, $\gamma$: the gyromagnetic factor. $\mathrm{H}$: the module of the RF or optical magnetic field. $g(\nu)$: the line shape function describing the spectral broadening of the absorption line.

$\sigma^{2}$ is a dimensionless factor representing the strength of the transition probalility, which depends on the relative orientation of the oscillating magnetic field with respect the crystal axis $z$. $\sigma^{2}$ is computed from the matrix element of the interaction hamiltonian, which depends on the total magnetic momentum and on the applied oscillating magnetic field. In our case where no static magnetic field is present, two different orientations have to be considered: $\mathrm{H}\perp z$ or $\mathrm{H}/\!/z$. We found that for each transition only one configuration gives noticeable value for $\sigma^{2}$. In the table \ref{Tab parametres}, we give the relevant $\sigma^{2}_{ij}$ parameters for each transition. The relative orientation of $\mathrm{H}$ with respect $z$ is indicated by the symbol ($\perp$) or ($/\!/$).

\begin{table}[h!]
\centering
\begin{tabular}{c c c c}
  \hline
 $O_{h}$  &  $\sigma_{1j}^{2}$ & $\sigma_{2j}^{2}$& $\sigma_{3j}^{2}$ \\
  \hline
             &      &     &     \\
$^{6}A_{1}(S)$ &   2.00 (0) [$\perp$] &  -  & -  \\

               &     3.27 (3) [$\perp$] &  1.25 (0) [$\perp$] & - \\

               &       &     &     \\

$^{4}T_{1}(G)$ &      1.84 (6) [$/\!/$] &    7.43 (5) [$\perp$] &    1.70 (6) [$\perp$] \\

               &      1.52 (5) [$\perp$] &    2.85 (6) [$/\!/$] &    9.63 (5) [$\perp$] \\

               &       8.53 (7) [$\perp$] &    1.88 (5) [$\perp$] &    1.33 (4) [$/\!/$] \\

               &      2.17 (5) [$\perp$] &    9.90 (5) [$/\!/$] &    6.19 (6) [$\perp$] \\

               &       4.08 (5) [$/\!/$] &    4.34 (6) [$\perp$] &    4.63 (7) [$\perp$] \\

               &       1.95 (5) [$/\!/$] &    8.90 (6) [$\perp$] &    3.64 (7) [$\perp$] \\

                 &     &     &     \\

$^{4}T_{2}(G)$ &        6.87 (4) [$/\!/$] &    2.08 (4) [$\perp$] &    2.74 (4) [$\perp$] \\

               &        5.92 (5) [$/\!/$] &    1.37 (7) [$\perp$] &    4.70 (4) [$\perp$] \\

               &       5.35 (6) [$\perp$] &    2.21 (4) [$/\!/$] &    1.61 (4) [$\perp$]  \\

               &       2.97 (5) [$\perp$] &    5.56 (4) [$\perp$] &    1.43 (4) [$/\!/$] \\

               &      1.71 (4) [$\perp$] &    5.12 (10) [$\perp$] &    4.06 (4) [$/\!/$] \\

               &       1.73 (4) [$\perp$] &    4.65 (4) [$/\!/$] &    1.36 (5) [$\perp$] \\

               &        &      &    \\

$^{4}A_{1}(G)$ &        3.43 (3) [$/\!/$] &    1.41 (4) [$\perp$] &    4.29 (4) [$\perp$]\\

               &        7.01 (4) [$\perp$] &    3.24 (5) [$/\!/$] &    1.05 (3) [$\perp$] \\
                &     &      &    \\

$^{4}E(G)$     &     \bf  1.40 (3) [$\perp$] &    \bf    1.80 (4) [$\perp$] &     \bf   8.32 (4) [$/\!/$] \\

               &       1.29 (4) [$\perp$] &    1.16 (3) [$\perp$] &    1.83 (3) [$/\!/$] \\

               &        8.88 (4) [$/\!/$] &    1.70 (3) [$\perp$] &    6.70 (4) [$\perp$]  \\

               &       2.67 (5) [$\perp$] &    4.62 (3) [$/\!/$] &    1.26 (3)  [$\perp$] \\

                &     &      &    \\

$^{4}T_{2}(D)$ &       5.55 (5) [$/\!/$] &    2.23 (5) [$\perp$] &    2.90 (5) [$\perp$] \\

               &       6.96 (6) [$\perp$] &    7.68 (6) [$/\!/$] &    5.84 (5) [$\perp$] \\

               &        4.50 (5) [$\perp$] &    4.93 (5) [$\perp$] &    1.62 (4) [$/\!/$] \\

               &        1.39 (6) [$/\!/$] &    8.22 (6) [$\perp$] &    1.01 (5) [$\perp$] \\

               &       7.80 (5) [$\perp$] &    2.42 (4) [$/\!/$] &    2.90 (5) [$\perp$] \\

               &        4.73 (5) [$/\!/$] &    9.33 (7) [$\perp$] &    8.77 (5) [$\perp$] \\

               &         &     &     \\

$^{4}E(D)$  &       9.67 (8) [$/\!/$] &    2.80 (7) [$\perp$] &    2.31 (8) [$\perp$]  \\

            &       9.12 (9) [$\perp$] &    5.53 (7) [$\perp$] &    7.05 (7) [$/\!/$]   \\

            &       2.29 (7) [$\perp$] &    1.11 (7) [$/\!/$] &    1.80 (7) [$\perp$]    \\

            &      2.23 (8) [$\perp$] &     7.54 (8) [$\perp$]  &    4.09 (7)  [$/\!/$] \\

                  &     &     &    \\

$^{4}T_{1}(P)$ &       1.27 (5) [$/\!/$] &    1.71 (7) [$\perp$] &    6.22 (7) [$\perp$] \\

               &       4.15 (6) [$/\!/$] &     5.79 (7) [$\perp$] &    2.58 (8) [$\perp$]  \\

               &      3.14 (5) [$/\!/$] &    1.31 (5) [$\perp$] &    8.33 (7)  [$\perp$]  \\

               &      4.22 (9) [$\perp$] &    6.69 (6) [$/\!/$] &    4.27 (5) [$\perp$] \\

               &        2.35 (6) [$/\!/$] &    2.19 (5) [$\perp$] &    5.23 (8) [$\perp$]  \\

               &     9.36 (7) [$/\!/$] &    1.61 (7) [$\perp$] &    1.59 (8)  [$\perp$]  \\

                &     &     &     \\

$^{4}A_{2}(F)$ &       2.10  (8) [$\perp$] &    6.58 (7) [$/\!/$] &    2.63 (7) [$\perp$] \\

               &       1.20 (7) [$/\!/$] &     6.33 (11) [$\perp$] &    2.90 (7) [$\perp$] \\
\hline
\end{tabular}
\caption{Strength of transition probability $\sigma^{2}$ for the $Fe^{3+}$  d-d transitions in $\alpha-Al_{2}O_{3}$.
The integer $(n)$ in brackets is for $\times 10^{-n}$).}
\label{Tab parametres}
\end{table}

\section{Proposal for a 31 GHz maser}

Laser sources emitting some 100 mW at 555 nm with a linewidth less than 0.1 nm are currently commercially available (see for ex. \cite{laser2012}). From the table \ref{Tab parametres} we selected the transition to the $^{4}E(G)$ excited level, at 18015 $cm^{-1}$ ($\approx 555$ nm). The transition probabilities for the $^{6}A_{1}(S)\longrightarrow$  $^{4}E(G)$ transition with $\mathrm{H}\perp z$ are: $\sigma^{2}_{1}$= 1.40$\times 10^{-3}$,  $\sigma^{2}_{2}$= 1.80$\times 10^{-4}$ and  $\sigma^{2}_{3}$ $\approx 0$. Thus if we submit the crystal to a narrowband 555 nm laser source, the $|1\rangle$ level will be more effectively pumped, and thus large population differences between $|1\rangle$ and the two other ground state levels can be obtained.

\subsection{Rate equations}
\begin{figure}[h!!!]
\centering
   \includegraphics[scale =0.4] {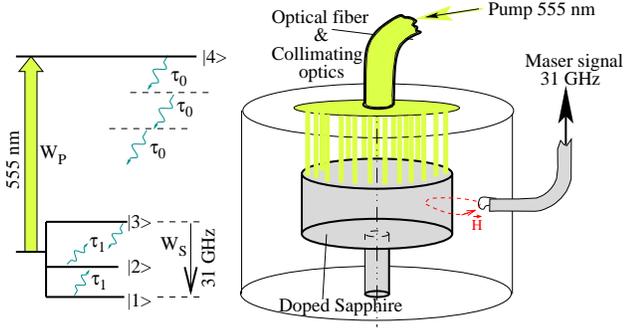}
\caption{Energy levels and schematic of the cryogenic optically pumped 31 GHz WhiGMO . \label{fig:31GHz-whigmo-principle}}
\end{figure}

Let'us consider a $Fe^{3+}$ doped $Al_{2}O_{3}$ crystal submitted to a 555 nm narrowband source resonant with the transition $^{6}A_{1}(S)\longrightarrow$  $^{4}E(G)$. The proposed system is schematized in figure \ref{fig:31GHz-whigmo-principle}. A cylindrical cryogenic sapphire resonator presenting a high-Q resonance mode at 31 GHz is submitted to an optical radiation at 555 nm. The crystal axis $z$ is colinear to the cylinder axis. The 31 GHz resonant mode is assumed to be quasi-transverse magnetic, i.e. the magnetic field lines in the equatorial plane of the cylinder are perpendicular to the $z$ axis. A small magnetic loop placed near the dielectric cylinder is used to derive the 31 GHz maser signal output. 

The excited levels decay rapidly to the ground state with a lifetime $\tau_{0} \ll1$ ms. All relaxation in the ground state are governed by the spin-lattice effect with a characteristic time $\tau_{1}\approx 10$ ms at 4 K.
It is also well established that the spin-to-spin relaxation is very fast with a characteristic time $\tau_{2}\leq 20$ ns. In this case a simple rate equations model dealing only with the level populations and neglecting coherences is sufficient to describe the interaction \cite{mrad2012}. Excited ions can possibly fall on intermediate levels, nevertheless as $\tau_{0} \ll\tau_{1}$ they relax rapidly to ground state. This allows us to only consider a four levels system, i.e. the three ground state levels and the $^{4}E(G)$ excited one, which is labelled $|4\rangle$. The ground state ions submitted to the optical radiation will be excited with an absorption rate $W_{P}$, which is evaluated from equation \ref{equ:taux-transition}. 
The optical pumping offers the possibility to get a population inversion between $|3\rangle$ and $|1\rangle$ and thus to obtain a maser effect at 31 GHz providing the resonator has been designed to present a high Q resonance at this frequency. The first 31 GHz emitted photons will stimulate others emissions with a rate $W_{S}$. If the optical pumping is effective enough to compensate for the resonator loss, a 31 GHz self-sustained oscillation will occur. To theoretically determine the conditions to get a 31 GHz maser signal we followed the method described in \cite{Benm2010} and validated in the case of 12 GHz WhiGMO. 
The stationary solutions of the rate equations are :
\begin{subequations}
\begin{equation}
\Delta n_{13}= \frac{27\Delta N_{13} + 6 (2 \Delta N_{13} - 3 \Delta N_{14}) \tau_{1} W_{P}}{9(3 + 4\tau_{1}W_{S}) + 4 (3+\tau_{1}W_{S})\tau_{1} W_{P}}
\label{eq11a}
\end{equation}
\begin{equation}
\Delta n_{14}= \frac{9(3\Delta N_{14}+2(2\Delta N_{14}-\Delta N_{13})\tau_{1}W_{S})}{9(3 + 4\tau_{1}W_{S}) + 4 (3+\tau_{1}W_{S})\tau_{1} W_{P}}
\label{eq11b}
\end{equation}
\end{subequations}

 $\Delta N_{ij}$ and $\Delta n_{ij}$ are the population difference between levels $|i\rangle$ and $|j\rangle$ at the thermal equilibrium and in presence of the pump and maser signal respectively. The pump and maser signal frequencies are denoted $\nu_{14}$ and $\nu_{13}$ respectively.

\subsection{Threshold pump power}
\label{sec:threshold-power}
In the absence of the maser signal, the threshold pump power is achieved when the population inversion takes place. The absorption rate $W_{P_{0}}$ corresponding to this situation is found by making $\Delta n_{13}$= 0 in the equation (\ref{eq11a}):
\begin{equation}
W_{P_{0}}= \frac{9}{2\tau_{1}} \frac{\Delta N_{13}}{3 \Delta N_{14} - 2 \Delta N_{13}}
\label{eq12}
\end{equation}
The total absorbed power is written as: 
\begin{align}
P_{0}&= h\nu_{14}W_{P_{0}}\Delta n_{14_{0}}NV_{eff} \label{eq13} \\
&= \frac{3}{2\tau_{1}}h\nu_{14}\Delta N_{13}NV_{eff} \nonumber
\end{align}
$N$ is the $Fe^{3+}$ ions concentration and $V_{eff}$ is the volume occupied by the active ions participating to the maser signal. $V_{eff}$ is the volume of the 31 GHz mode in the sapphire resonator (see section \ref{sec:resonator-design}).
Here, we assume the optical radiation only lights the resonator part where the 31 GHz mode is concentrated. $P_{0}$ is thus the minimal optical power required to get maser action in optimized conditions.

\subsection{Maximum maser signal power}
\label{sec:max-power}

$\Delta n_{13}$ and thus the 31 GHz signal power increase with the pump signal power system until saturation of the optical transition occurs, i.e $n_{4} \simeq n_{1}$. Assuming we already overtook this situation, i.e. $W_{P} \rightarrow \infty$, equations (11) simplify and the maximum maser power is derived as:

\begin{align}
P_{31_{max}}&=h\nu_{13} \ Max[W_{S}\Delta n_{13}] N V_{eff} \label{14}\\
&=\frac{3}{2\tau_{1}}h\nu_{13}(3\Delta N_{14}-2\Delta N_{13})N V_{eff} \nonumber
\end{align}

$P_{31_{max}}$ is the total power generated by the active ions inside the resonator. The coupling loop derive only a fraction of this power, which depends on the coupling coefficient $\beta$ related to the loop area and to its position with respect to the sapphire resonator. For a critical coupling , i.e. $\beta =1$,  the maser output power is one half of $P_{31_{max}}$. The resonator Q-factor is also impacted by the loading to the external circuit and for  $\beta =1$, the loaded Q-factor is one half of the unloaded one. 

\subsection{Resonator design}
\label{sec:resonator-design}

Our objective is to use the $Fe^{3+}$ ions present as contaminant in a high purity sapphire monocrystal. Typical effective ions concentration in such a type a crystal is well below 1 ppm. Thus the 31 GHz resonator has to present a very high quality factor, otherwise the 31 GHz stimulated emissions wont be sufficient to compensate for the resonator loss. This can be accomplished by using a whispering gallery mode, which Q-factor is only limited by the sapphire losses. 
Quasi--TM Whispering gallery modes are characterized by
three integers $ {m, n, l}$\ representing the electromagnetic
field component variations along the azimutal, radial and axial
directions respectively \cite{jiao1987}. In the following we consider
only the resonant modes with low radial and axial variations ,\
i.e. those corresponding to $n=l=0$. 
At low temperature, i.e. near the liquid helium temperature, the whispering gallery mode Q-factor can be as high as 1 billion providing the mode order is sufficiently high, i.e.  $m >15$ \cite{bourg(2004)}. The WhiGMO principle has been demonstrated at 12 GHz, i.e. the frequency of the $|1\rangle \rightarrow |2\rangle$ transition. In this case, the population inversion results from a pumping at 31 GHz. 
The resonator has a diameter $2R=50$ mm and a thichness $h=30$ mm. The quasi-TM Whispering Gallery mode $WGH_{17,0,0}$ is used to support the maser signal \cite{bourg(2006)}. \\

For the whispering gallery mode the electromagnetic fields are concentrated near curved air-dielectric interface, between the resonator radius $R$ and a caustic of radius $R_{C}$. The effective volume $V_{eff} = \pi h \left (R^{2}-R^{2}_{C} \right )$ occupied by the mode is a little mode dependant as $R_{C}$ is given by:

\begin{equation}
R_{C} = \frac{m} {\sqrt{ \epsilon_{r} \left ( \dfrac{\nu_{13}}{c} \right )^{2}-  \left ( {\dfrac{\pi}{2h}} \right
 ) ^{2} }} \label{equ:caustic}
\end{equation}
where $\epsilon_{r} \approx 9.4$ is the sapphire permittivity is the transverse direction with respect $z$. $c$ the light velocity  in vacuum.\\

To design a 31 GHz resonator we keep the same ratio diameter/thickness, which is near the optimal value \cite{krup1996}. Thus to design the cryogenic resonator only means choosing the azimuthal number for the mode supporting the 31 GHz signal. $m$ has to be higher than $15$ to prevent radiation losses which limit the Q-factor and should not be to high otherwise the coupling with the output probe will be difficult to adjust. The table \ref{tab:resonator-design} gives for different values of $m$, the resonator diameter $2R$, the threshold optical power $P_{0}$ and the maximum maser power emitted by the ions. Here, the $Fe^{3+}$ concentration is assumed to be $0.02$ ppm, i.e. the same value than the 12 GHz WhiGMO.

\begin{table}[h!]
\centering
\begin{tabular}{|c|c|c|c|}
  \hline
$m$ 		& $2R$ 		& $P_{0}$ 	& $P_{31_{max}}$ \\
 		& (mm)		& (mW)		& ($\mu$W) \\	
  \hline 
   \hline 
17 		&  19.22		& 1.17  		& 1.02 \\ 
  \hline 
19 		&  21.13		& 1.42 		& 1.24 \\ 
  \hline 
21 		&   23.05		& 1.71  		& 1.49 \\
\hline 
25 		&   26.88 		& 2.35 		&  2.05  \\
\hline 
35 		&   36.46		& 4.40 		& 3.85  \\
 \hline 
49 		& 50.00  		&8.41  		& 7.36  \\
\hline
\end{tabular}
\caption{Different resonator designs to support the 31 GHz signal. The pump threshold power $P_{0}$ and the maximum maser power $P_{31_{max}}$ are given for few azimutal numbers as well as the resonator diameter assuming a constant ratio $\frac{2R}{h}=\frac{5}{3}$.}
\label{tab:resonator-design}
\end{table}

\end{document}